\documentclass [a4paper,fleqn, 12pt]{article}
\usepackage{graphicx}
\usepackage[small]{subfigure,epsfig}

\usepackage {amsmath} \usepackage{amssymb} \usepackage{cite}

\usepackage{cite}

\newcommand{\PDfrac}[3][{}]{ \frac{\partial^{#1}#2}{\partial#3^{#1}} }

\begin{document}

\title{Equation for three--dimensional nonlinear waves in liquid with gas bubbles}
\author{Nikolay A. Kudryashov and Dmitry I. Sinelshchikov}
\date{Department of Applied Mathematics, National Research Nuclear University
MEPHI, 31 Kashirskoe Shosse, 115409 Moscow, Russian Federation}
\maketitle

\begin{abstract}
Nonlinear waves in a liquid containing gas bubbles are considered in the three--dimensional case. Nonlinear evolution equation is given for description of long nonlinear pressure waves. It is shown that in the general case  the equation is not integrable. Some exact solutions for the nonlinear evolution equation are presented. Application of the Hirota method is illustrated for finding multi--soliton solutions for the nonintegrable evolution equation in the three-dimensional case. The stability of the one--dimensional solitary waves is investigated. It is shown that the one--dimensional solitary waves are stable to transverse perturbations.

\end{abstract}



\section{Introduction.}

In the nature, technology, and medicine we often can observe the medium consisting of a liquid with gas bubbles \cite{Nakor'akov, Nigmatulin, Goldberg}. One of the important problem for investigation in bubbly liquids is propagation of waves. Wave processes in bubble--liquid mixtures were considered in many papers. Taking into account the linear approach we have the well known problem that does not differ in essence from many other problems of the linear wave mechanics.

However a lot of new possibilities appear when we consider the nonlinear processes. The influence of nonlinearity on waves propagation in a liquid with gas bubbles was first studied in  \cite{Wijngaarden,Wijngaarden1972,Nakor'akov_1972}. It was shown that long weakly nonlinear waves in a bubbly liquid can be described by the Korteweg--de Vries equation. Authors also took into consideration the viscosity of liquid and obtained the Korteweg--de Vries--Burgers equation for the description of nonlinear waves. However in \cite{Wijngaarden,Wijngaarden1972,Nakor'akov_1972} only one--dimensional weakly nonlinear waves were considered.

Recently \cite{Kudryashov2010, Kudryashov2010A} we have studied the influence of heat--transfer between a liquid and a gas in bubbles on propagation of weakly nonlinear waves in bubble--liquid mixtures. As a result we have obtained nonlinear evolution equations of the second, third and fourth order for the description of wave processes in a one--dimensional bubble--liquid mixture.

The real nonlinear waves in a liquid with gas bubbles are two- and three-dimensional. So we need to use multidimensional mathematical models for studying nonlinear waves in a bubbly liquid. Some problems of pressure wave distributions in a bubble--liquid mixture in the two--dimensional case were considered numerically in \cite{Shagapov2002, Shagapov2006, Kudryashov2009_MM}.

However the calculations of wave processes presented in these works are based on cumbersome mathematical models and do not allow us to reveal basic parameters of nonlinear wave processes in a bubbly liquid. Besides we cannot find any exact solutions for these models and consequently cannot effectively analyze correctness of numerical modeling.

Thus it is important to consider long weakly nonlinear waves in a liquid with gas bubbles in the three-dimensional case. In this case we can obtain evolution equation for the nonlinear waves in a bubbly liquid. The evolution equation can be studied both analytically and numerically. Also analyzing the evolution equation we can determine nature of physical characteristics influence on nonlinear waves.

The aim of this work is to study long weakly nonlinear waves in a liquid containing gas bubbles in the three-dimensional case. In section 2 we present governed system of equations for nonlinear waves in a bubbly liquid. In section 3 with the help of the reductive perturbation method we obtain the evolution equation for long weakly nonlinear weakly three-dimensional waves in a liquid with gas bubbles. In section 4 we show that this equation is not integrable in the general case. In section 5 we obtain some exact solutions for the evolution equation. We present stability analysis of one-dimensional solitary waves to transverse perturbations in section 5 as well.

\section{System of equations for the description of wave processes in liquid with gas bubbles.}

Let us suppose that a liquid containing gas bubbles is a homogeneous medium, which has an average pressure and a temperature \cite{Nakor'akov,Nigmatulin}.
We assume that the liquid is incompressible and bubbles are spherical. We do not take into consideration formation, destruction, interaction and coalescence of bubbles. Let us assume that all gas bubbles have the same size and the amount of bubbles in the mass unit is constant $N$. We suppose that the gas in bubbles is ideal and pressure of gas in bubbles is governed by polytrophic law. We shall consider the viscosity of the liquid only in the interphase boundary. Under these assumptions the following closed system of equations may be used for the description of waves processes in the liquid containing gas bubbles \cite{Nakor'akov,Nigmatulin}
\begin{equation}
\PDfrac{\rho}{\tau}+\mbox{div}\rho\,\mathbf{u}=0,
\label{pre_main_main_dim_system_1}
\end{equation}
\begin{equation}
\rho\left(\PDfrac{\mathbf{u}}{\tau}+\mathbf{u}\nabla\mathbf{u}\right)+\nabla P=0,
\label{pre_main_main_dim_system_2}
\end{equation}
\begin{equation}
\rho_{l}\left(R R_{\tau\tau}+\frac{3}{2} R_{\tau}^{2}+\frac{4\nu}{3R}R_{\tau}\right)=P_{g}-P-\frac{2\sigma}{R},
\label{pre_main_main_dim_system_3}
\end{equation}
\begin{equation}
P_{g}=P_{g,0}\left(\frac{R_{0}}{R}\right)^{3\,n},
\label{pre_main_main_dim_system_4}
\end{equation}
\begin{equation}
\rho=(1-\phi)\,\rho_{l}+\phi\,\rho_{g},
\label{pre_main_main_dim_system_5}
\end{equation}
\begin{equation}
\phi=V\,\rho, \quad V=\frac{4}{3}\,\pi\,R^{3}\,N
\label{pre_main_main_dim_system_6}
\end{equation}
where $\xi,\eta,\zeta$ are cartesian coordinates, $\rho(\xi,\eta,\zeta,\tau)$ is the density of the bubble--liquid mixture, $P(\xi,\eta,\zeta,\tau)$ is the pressure of the mixture, $\mathbf{u}(\xi,\eta,\zeta,\tau)=(u^{(1)}(\xi,\eta,\zeta,\tau),u^{(2)}(\xi,\eta,\zeta,\tau),
u^{(3)}(\xi,\eta,\zeta,\tau))^{T}$ is the velocity of the mixture, $R=R(\xi,\eta,\zeta,\tau)$ is bubbles radius, $\rho_{l}, \rho_{g}(\xi,\eta,\zeta,\tau)$ are densities of the liquid and the gas respectively, $P_{g}(\xi,\eta,\zeta,\tau)$ is the pressure of the gas, $P_{g,0}$ and $R_{0}$ are the pressure of the gas and the radius of bubbles in the unperturbed state, $\sigma$ is the surface tension, $\nu$ is the kinematic viscosity of the liquid, $\phi$ is the volume gas content, $V$ is the gas volume in the unit mass of the mixture, $n$ is the polytrophic exponent and the nabla operator is given by $\nabla=(\partial/\partial\xi,\partial/\partial\eta,\partial/\partial\zeta)$.

We shall assume that in the unperturbed state the pressure and the density of the bubble--liquid mixture are constants, bubbles have all the same radius and uniformly distributed in the liquid.

Assuming that the mean inter-bubble distance $d$ is much greater than the bubble radius, we do not take into account bubbles interaction. We suppose that the characteristic wavelength $L$ is greater than linear scale of a region containing many bubbles \cite{Wijngaarden,Nakor'akov_1972,Nakor'akov,Nigmatulin}. Thus we have the following relation between characteristic length scales in the bubbly liquid
\begin{equation}
R_{0} \ll d \ll L
\label{scales}
\end{equation}

Assuming that the volume gas content is small $\phi \ll 1$  we have from \eqref{pre_main_main_dim_system_5} and \eqref{pre_main_main_dim_system_6}
\begin{equation}
\rho=\frac{\rho_{l}}{1+\rho_{l}\,V}
\label{eq: density}
\end{equation}

Using \eqref{eq: density} system \eqref{pre_main_main_dim_system_1}--\eqref{pre_main_main_dim_system_6} can be written as
\begin{equation}
\PDfrac{\rho}{\tau}+\mbox{div}\rho\,\mathbf{u}=0,
\label{main_dim_system_1}
\end{equation}
\begin{equation}
\rho\left(\PDfrac{\mathbf{u}}{\tau}+\mathbf{u}\nabla\mathbf{u}\right)+\nabla P=0,
\label{main_dim_system_2}
\end{equation}
\begin{equation}
\rho_{l}\left(R R_{\tau\tau}+\frac{3}{2} R_{\tau}^{2}+\frac{4\nu}{3R}R_{\tau}\right)=P_{g,0}\left(\frac{R_{0}}{R}\right)^{3\,n}-P-
\frac{2\sigma}{R},
\label{main_dim_system_3}
\end{equation}
\begin{equation}
\rho=\frac{\rho_{l}}{1+\rho_{l}\,V}, \quad V=\frac{4}{3}\,\pi\,R^{3}\,N.
\label{main_dim_system_4}
\end{equation}

Let us suppose that deviation of the density from the unperturbed density is small in comparison with the last
\begin{equation}
  \begin{gathered}
 \rho(\xi,\eta,\zeta,\tau) = \rho_{0} + \tilde{\rho}(\xi,\eta,\zeta,\tau),\:\   \quad
  \rho_{0}=\mbox{const},\quad  ||\tilde{\rho}||<<\rho_{0},
  \label{eq: rel1}
\end{gathered}
\end{equation}
where $\rho_{0}$ is the density of the bubble--liquid mixture in the unperturbed state.

Taking into account \eqref{eq: rel1} from \eqref{main_dim_system_4} with the accuracy up to terms of order $\tilde{\rho}^{2}$ we have
\begin{equation}
  \begin{gathered}
R=R_{0}-\mu\tilde{\rho}+\mu_{1}\tilde{\rho}^{2},  \vspace{0.1cm} \\
R_{0}^{3}=\frac{3}{4\pi N}\left(\frac{1}{\rho_{0}}-\frac{1}{\rho_{l}}\right), \quad \mu=\frac{R_{0}}{3\,\rho_{0}^{2}\,V_{0}}, \quad
\mu_{1}=\frac{R_{0}(3\,\rho_{0}V_{0}-1)}{9\,\rho_{0}^{4}\,V_{0}^{2}},\vspace{0.1cm}  \\
V_{0}=\frac{4}{3}\pi N R_{0}^{3}
  \label{eq:R_equation}
  \end{gathered}
\end{equation}

Substituting \eqref{eq:R_equation} into \eqref{main_dim_system_1}--\eqref{main_dim_system_4} and taking into account \eqref{eq: rel1} we obtain
\begin{equation}
  \begin{gathered}
\PDfrac{\tilde{\rho}}{\tau}+\rho_{0}\mbox{div}\mathbf{u}+\mbox{div}\tilde{\rho}\mathbf{u}=0, \hfill \vspace{0.2cm} \cr
(\rho_{0}+\tilde{\rho})\left(\PDfrac{\mathbf{u}}{\tau}+\mathbf{u}\nabla\mathbf{u}\right)+\nabla P=0,\hfill \vspace{0.2cm} \\
P=P_{0}-\frac{2\sigma}{R_{0}}+\left(\frac{3n\,\mu\,P_{0}}{R_{0}}-\frac{2\sigma\mu}
{R_{0}^{2}}\right)\,\tilde{\rho}
+\hfill \vspace{0.1cm} \\
+\left(\frac{3n[(3n+1)\,\mu^{2}-2\,\mu_{1}\,R_{0}]P_{0}}{2\,R_{0}^{2}}-\frac{\sigma
(\mu^{2}-\mu_{1}R_{0})}{R_{0}^{3}}\right)\,\tilde{\rho}^{2}+\hfill \vspace{0.1cm}\\+
\rho_{l}\,\mu\,R_{0}\tilde{\rho}_{\tau\tau}-\rho_{l}(2\mu_{1}\,R_{0}+\mu^{2})
\tilde{\rho}\,\tilde{\rho}_{\tau\tau}-
\rho_{l}\left(2\mu_{1}\,R_{0}+\frac{3\mu^{2}}{2}\right)\,\tilde{\rho}_{\tau}^{2}+\hfill \vspace{0.1cm} \\
+\frac{4\nu\mu\,\rho_{l}}{3R_{0}}\,\tilde{\rho}_{\tau}+
\frac{(4\nu\mu^{2}-8\nu\mu_{1}R_{0})\rho_{l}}{3R_{0}^{2}}\,\tilde{\rho}\,
\tilde{\rho}_{\tau},\hfill
\label{dim_system}
  \end{gathered}
\end{equation}

where we denote by $P_{0}$ the pressure of gas in bubbles in the unperturbed state.

Linearizing system \eqref{dim_system} and assuming that $P$ is proportional to $\tilde{\rho}$ we obtain the following equation
\begin{equation}
\tilde{\rho}_{\tau\tau}=c_{0}^{2} \, \triangle \tilde{\rho}, \quad c_{0}^{2}=\frac{3n\mu P_{0}}{R_{0}}-\frac{2\sigma\mu}{R_{0}^{2}},
\label{wave_equation}
\end{equation}
where $\triangle$ is the Laplace operator.
From \eqref{wave_equation} we can see that speed of waves in the linear case is $c_{0}$. Further we shall use $c_{0}$ for introducing dimensionless variables.

Let us introduce the dimensionless variables
\begin{equation}
  \begin{gathered}
\xi = L\, \xi^{'}, \quad \eta=L\, \eta^{'},\quad \zeta=L\, \zeta^{'}  \quad   \tau= \frac{ L }{ c_{0} }\, \tau', \quad  \mathbf{u} = c_{0}\, \mathbf{u}^{'}, \\
  \tilde{\rho}=\rho_{0} \tilde{\rho}^{'}, \quad P = P_0\, P'+P_0-\frac{2\sigma}{R_{0}},
  \label{eq: non-dim_subst}
  \end{gathered}
\end{equation}
where $P_{0}-2\sigma/R_{0}$ is the pressure of the mixture in the unperturbed state. The quantities $L$ and $\tau_{*}=L/c_{0}$ are characteristic length scale and characteristic time on which  parameters of the bubbly liquid change \cite{Nigmatulin}.

Using the dimensionless variables we can reduce \eqref{dim_system} to the following system of equations (the primes are omitted)
\begin{equation}
  \begin{gathered}
\PDfrac{\tilde{\rho}}{\tau}+\mbox{div}\mathbf{u}+\mbox{div}\tilde{\rho}\,\mathbf{u}=0, \hfill \vspace{0.1cm} \cr
(1+\tilde{\rho})\left(\PDfrac{\mathbf{u}}{\tau}+\mathbf{u}\nabla\mathbf{u}\right)+
\frac{1}{\alpha}\nabla P=0,\hfill \vspace{0.1cm} \\
P=\alpha\tilde{\rho}+\alpha_{1}\tilde{\rho}^{2}+
\beta\tilde{\rho}_{\tau\tau}-(\beta_{1}+\beta_{2})\tilde{\rho}\,\tilde{\rho}_{\tau\tau}-
(\beta_{1}+\frac{3}{2}\beta_{2})\tilde{\rho}_{\tau}^{2}+\hfill \\+
\varkappa\tilde{\rho}_{\tau}+\varkappa_{1}\tilde{\rho}\tilde{\rho}_{\tau} \hfill
\label{main_nondim_system}
  \end{gathered}
\end{equation}

\begin{equation}
  \begin{gathered}
    \alpha = \frac{3n\mu\rho_{0}}{R_{0}}-\frac{2\sigma\mu\rho_{0}}{R_{0}^{2}P_{0}},\quad
    \beta=\frac{\rho_{l}\,\mu R_{0} c_{0}^{2} \rho_{0}}{P_{0}\,L^{2}}, \\
    \alpha_{1}=\frac{3n[(3n+1)\,\mu^{2}-2\,\mu_{1}\,R_{0}]\rho_{0}^{2}}{2\,R_{0}^{2}}
    -\frac{\sigma(\mu^{2}-\mu_{1}R_{0})\rho_{0}^{2}}{P_{0}R_{0}^{3}} \cr
    \beta_{1} = \frac{2\rho_{l}\,\mu_{1} R_{0} c_{0}^{2} \rho_{0}^{2}}{P_{0}\,L^{2}}, \quad
    \beta_{2} = \frac{\rho_{l}\,\mu^{2} \rho_{0}^{2} c_{0}^{2}}{P_{0}\,L^{2}}, \\
    \varkappa=\frac{4\nu\mu\,\rho_{l}\rho_{0}c_{0}}{3R_{0}P_{0}L},\quad
    \varkappa_{1}=\frac{(4\nu\mu^{2}-8\nu\mu_{1}R_{0})\rho_{l}\rho_{0}^{2}c_{0}}
    {3R_{0}^{2}P_{0}L}
  \label{eq: non-dim_parameters}
  \end{gathered}
\end{equation}

Further we shall analyze this system of equations with the help of the reductive perturbation method. From \eqref{main_nondim_system} we shall obtain the nonlinear evolution equation for long weakly nonlinear weakly three-dimensional waves in the liquid with gas bubbles.


\section{Nonlinear Evolution Equation for long waves in the liquid with gas bubbles in the three-dimensional case.}

Let us consider transverse variations of the long waves governed by \eqref{main_nondim_system}. We suppose that waves propagate along  the $x$-axis with allowance to transverse evolution in $y$ and $z$ directions. To obtain the nonlinear evolution equation for the long waves we shall use the reductive perturbation method, which was introduced in \cite{Taniuti1966,Gardner1969}. The application of the reductive perturbation method to multidimensional problems, for example, were considered in \cite{Kako1976,Taniuti1991,Parkes1997,Parkes1999}. For the modern review on the reductive perturbation method see \cite{Leblond2008}.

In accordance with reductive perturbation method we need to introduce scale transformations of the independent variables. To this end we use the dispersion relation for the linearized system \eqref{main_nondim_system}, which in the case of $\varkappa=0$ is the following
\begin{equation}
k_{\xi}^{2}+k_{\eta}^{2}+k_{\zeta}^{2}=\frac{\omega^{2}}{1-\frac{\beta}{\alpha}\omega^{2}}
\label{dispersion_relation}
\end{equation}
where $k_{\xi}$, $k_{\eta}$ and $k_{\zeta}$ are wave numbers and $\omega$ is a frequency.

As we consider weakly three-dimensional waves we neglect wave numbers $k_{\eta}$ and $k_{\zeta}$ in \eqref{dispersion_relation}. Assuming that $\omega^{2} \ll 1$ form \eqref{dispersion_relation} we obtain
\begin{equation}
k_{\xi} \simeq \omega + \frac{\beta}{2\alpha} \omega^{3}
\end{equation}
Assuming that $\omega^{2}=\varepsilon \tilde{\mu}^{2}$ we see that phase of plane wave in the $x$-direction has the form
\begin{equation}
k_{\xi}\,\xi-\omega\,\tau=\tilde{\mu} \left(\varepsilon^{1/2}(\xi-\tau)+\frac{\beta}{2\alpha}\tilde{\mu}^{2}\varepsilon^{3/2} \xi\right)
\label{dispersion_relation_1}
\end{equation}

Using \eqref{dispersion_relation_1} we can introduce 'slow' variables in \eqref{main_nondim_system} as follows
\begin{equation}
  x = \varepsilon^{1/2}(\xi-\tau), \quad t= \varepsilon^{3/2}\, \tau ,\quad
  y=\varepsilon\,\delta\,\eta, \quad z=\varepsilon\,\gamma\,\zeta \quad \varepsilon \ll 1.
  \label{eq: rescaling_coordinates}
\end{equation}
Then we have the following relations
\begin{equation}\label{eq100}
  \frac{\partial}{\partial \xi} = \varepsilon^{1/2} \frac{\partial}{\partial x},\quad
  \frac{\partial}{\partial \tau} = \varepsilon^{3/2} \frac{\partial}{\partial t}
  - \varepsilon^{1/2} \frac{\partial}{\partial x}, \quad
  \frac{\partial}{\partial \eta} = \varepsilon \delta \frac{\partial}{\partial y}, \quad
  \frac{\partial}{\partial \zeta} = \varepsilon \gamma \frac{\partial}{\partial z}
\end{equation}

In \eqref{eq: rescaling_coordinates} we suppose that transverse variables $y$ and $z$ are 'slower' than $x$ as far as we consider transverse perturbation of the wave propagating in the $x$ direction.

We consider the long waves in the liquid with gas bubbles. Thus the characteristic wave length is much greater than the mean inter-bubble distance  and the radius of bubbles in the unperturbed state (see \eqref{scales}).
In this case we can interpret small parameter $\varepsilon$ as $\varepsilon=\frac{R_{0}}{L}\ll 1$. Also we assume that $\delta$ and $\gamma$ are parameters which characterize perturbations in $y$ and $z$ directions.

Substituting \eqref{eq: rescaling_coordinates} in \eqref{main_nondim_system} and dividing on $\varepsilon^{1/2}$  in
first three equations we have the system of equations in the form
\begin{equation}
  \begin{gathered}
\varepsilon\tilde{\rho}_{t}-\tilde{\rho}_{x}+u^{(1)}_{x}+\varepsilon^{1/2}\delta\,u^{(2)}_{y}+
\varepsilon^{1/2}\gamma u^{(3)}_{z}
+(\tilde{\rho}u^{(1)})_{x}+\varepsilon^{1/2}\delta(\tilde{\rho}u^{(2)})_{y}+\hfill \\+\varepsilon^{1/2}
\gamma(\tilde{\rho}u^{(3)})_{z}=0 \hfill ,
\label{eq: general_rescaled_system_1}
  \end{gathered}
\end{equation}
\begin{equation}
  \begin{gathered}
 (1+\tilde{\rho})(\varepsilon u^{(1)}_{t}-u^{(1)}_{x}+u^{(1)}\,u^{(1)}_{x}+\varepsilon^{1/2}\delta\,u^{(2)}\,u^{(1)}_{y}+\varepsilon^{1/2}
\gamma\,u^{(3)}\,u^{(1)}_{z})+\hfill \\+\frac{1}{\alpha}\,P_{x}=0 \hfill ,
\label{eq: general_rescaled_system_2}
  \end{gathered}
\end{equation}
\begin{equation}
  \begin{gathered}
 (1+\tilde{\rho})(\varepsilon\,u^{(2)}_{t}-u^{(2)}_{x}+u^{(1)}\,u^{(2)}_{x}+\varepsilon^{1/2}\delta u^{(2)}\,u^{(2)}_{y}+\varepsilon^{1/2}\gamma u^{(3)}\,u^{(2)}_{z})+\hfill \\+\varepsilon^{1/2}\frac{\delta}{\alpha}P_{y}=0 \hfill ,
\label{eq: general_rescaled_system_3}
  \end{gathered}
\end{equation}
\begin{equation}
  \begin{gathered}
 (1+\tilde{\rho})(\varepsilon\,u^{(3)}_{t}-u^{(3)}_{x}+u^{(1)}\,u^{(3)}_{x}+\varepsilon^{1/2}\delta u^{(2)}\,u^{(3)}_{y}+\varepsilon^{1/2}\gamma u^{(3)}\,u^{(3)}_{z})+\hfill \\+\varepsilon^{1/2} \frac{\gamma }{\alpha}P_{z}=0,\hfill 
\label{eq: general_rescaled_system_4}
  \end{gathered}
\end{equation}
\begin{equation}
  \begin{gathered}
P=\alpha\,\tilde{\rho}+\alpha_{1}\,\tilde{\rho}^{2}+\varepsilon^{3}\,\beta\,
\tilde{\rho}_{tt}+\varepsilon^{2}\,\beta\,\tilde{\rho}_{tx}+\varepsilon\,\beta\,
\tilde{\rho}_{xx}-
\varepsilon^{3}\,(\beta_{1}+\beta_{2})\tilde{\rho}\,\tilde{\rho}_{tt}+  \hfill \\
+\varepsilon^{2}\,2(\beta_{1}+\beta_{2})\tilde{\rho}\,\tilde{\rho}_{tx}-
\varepsilon\,(\beta_{1}+\beta_{2})\tilde{\rho}\,\tilde{\rho}_{xx}-\varepsilon^{3}
\,(\beta_{1}+\frac{3}{2}\beta_{2})\tilde{\rho}_{t}^{2}+\hfill \\+
\varepsilon^{2}\,(\beta_{1}+\frac{3}{2}\beta_{2})\tilde{\rho}_{x}\tilde{\rho}_{t}
  -\varepsilon\,(\beta_{1}+\frac{3}{2}\beta_{2})\tilde{\rho}_{x}^{2}
+\varepsilon^{3/2}\varkappa\,\tilde{\rho}_{t}-\varepsilon^{1/2}\varkappa\,
\tilde{\rho}_{x}
+\hfill \\+\varepsilon^{3/2}\varkappa_{1}\,\tilde{\rho}\,\tilde{\rho}_{t}-
\varepsilon^{1/2}\varkappa\,\tilde{\rho}\,\tilde{\rho}_{x}
\label{eq: general_rescaled_system_5}
  \end{gathered}
\end{equation}
Let us look for the solution of \eqref{eq: general_rescaled_system_1}--\eqref{eq: general_rescaled_system_5} in the form
\begin{equation}
  \begin{gathered}
    u^{(1)} = \varepsilon u^{(1)}_1 + \varepsilon^{2} u^{(1)}_2 +  \ldots, \qquad
    u^{(2)}= \varepsilon^{3/2} u^{(2)}_1 + \varepsilon^{5/2} u^{(2)}_2 +  \ldots, \hfill  \\
    u^{(3)}= \varepsilon^{3/2} u^{(3)}_1 + \varepsilon^{5/2} u^{(3)}_2 +  \ldots, \quad
    \tilde{\rho}  =\varepsilon \tilde{\rho} _1 + \varepsilon^{2} \tilde{\rho}_2 +\ldots, \hfill \\
     P = \varepsilon P_1  + \varepsilon^{2} P_2   +\ldots \hfill
  \label{eq: asymptotic_expansion1}
  \end{gathered}
\end{equation}

Substituting \eqref{eq: asymptotic_expansion1} into \eqref{eq: general_rescaled_system_1}--\eqref{eq: general_rescaled_system_5} and equating expressions at $\varepsilon$ we have
\begin{equation}
    -\tilde{\rho}_{1 x} + u^{(1)}_{1 x} = 0\, ,\quad
    - u^{(1)}_{1 x} + \frac{1}{\alpha}\, P_{1 x} = 0\, ,\quad
     P_{1} = \alpha \, \tilde{\rho}_{1}.
\end{equation}

We obtain solutions for the last equations in the form
\begin{equation}
  \begin{gathered}
    u^{(1)}_{1}(x,y,z,t) =\tilde{\rho}_{1}(x,y,z,t)\, +\, \tilde{u}^{(1)}_{1}(y,z,t)\, \hfill \\
    P_{1}(x,y,z,t) =\alpha \, \tilde{\rho}_{1}(x,y,z,t) \hfill 
  \label{eq: null_approximation_relations1}
  \end{gathered}
\end{equation}
where $\tilde{u}^{(1)}_{1}(y,z,t)$ is an arbitrary function. Without loss of generality we assume that $\tilde{u}^{(1)}_{1} (y,z,t)=0$.

Let us substitute \eqref{eq: asymptotic_expansion1} into
\eqref{eq: general_rescaled_system_1}--\eqref{eq: general_rescaled_system_5}. Equating expressions at $\varepsilon^{2}$ in \eqref{eq: general_rescaled_system_1}, \eqref{eq: general_rescaled_system_2}, \eqref{eq: general_rescaled_system_5}  and at $\varepsilon^{3/2}$ in \eqref{eq: general_rescaled_system_3}, \eqref{eq: general_rescaled_system_4} we obtain the following system of equations in the case of $\varkappa=O(\varepsilon^{1/2})$
\begin{equation}
  \begin{gathered}
  \tilde{\rho}_{1t}-\tilde{\rho}_{2x}+u^{(1)}_{2x}+\delta u^{(2)}_{1y}+\gamma u^{(3)}_{1z}+(\tilde{\rho}_{1}\,u^{(1)}_{1})_{x}=0, \hfill \\
  u^{(1)}_{1t}-u^{(1)}_{2x}+u^{(1)}_{1}\,u^{(1)}_{1x}-\tilde{\rho}_{1}\,u^{(1)}_{1x}+\frac{1}{\alpha}\,P_{2x}=0, \hfill \\
  -u^{(2)}_{1x}+\frac{\delta}{\alpha}\,P_{1y}=0, \hfill \\
  -u^{(3)}_{1x}+\frac{\gamma}{\alpha}\,P_{1z}=0, \hfill \\
  P_{2}=\alpha\,\tilde{\rho}_{2}+\alpha_{1}\,\tilde{\rho}_{1}^{2}+\beta\,
  \tilde{\rho}_{1xx}-\varkappa\,\tilde{\rho}_{1x}. \hfill
  \label{eq: first_approximation_relations}
  \end{gathered}
\end{equation}
Using \eqref{eq: null_approximation_relations1} we find from \eqref{eq: first_approximation_relations}
\begin{equation}
  \begin{gathered}
  \tilde{\rho}_{1t}+\left(1+\frac{\alpha_{1}}{\alpha}\right)\tilde{\rho}_{1}
  \tilde{\rho}_{1x}+\frac{\beta}{2\alpha}\tilde{\rho}_{1xxx}
  -\frac{\varkappa}{2\alpha}\tilde{\rho}_{1xx}+\frac{\delta}{2}\,u^{(2)}_{1y}+
  \frac{\gamma}{2}\,u^{(3)}_{1z}=0, \hfill \\
  u^{(2)}_{1x}=\delta \tilde{\rho}_{1y}, \hfill \\
  u^{(3)}_{1x}=\gamma \tilde{\rho}_{1z} \hfill
  \label{pre_KP}
  \end{gathered}
\end{equation}

We can write \eqref{pre_KP} as the equation
\begin{equation}
\left(\tilde{\rho}_{1t}+\left(1+\frac{\alpha_{1}}{\alpha}\right)\tilde{\rho}_{1}
\tilde{\rho}_{1x}+
\frac{\beta}{2\alpha}\tilde{\rho}_{1xxx}-\frac{\varkappa}{2\alpha}
\tilde{\rho}_{1xx}\right)_{x}+
\frac{\delta^{2}}{2}\,\tilde{\rho}_{1yy}+\frac{\gamma^{2}}{2}\,\tilde{\rho}_{1zz}=0
\label{KP}
\end{equation}
Equation \eqref{KP} is the three-dimensional counterpart of the Korteweg--de Vries--Burgers equation for description of nonlinear waves in the liquid with gas bubbles.

In the case of $\gamma=0$ (i.e. two-dimensional waves) and $\varkappa=0$ (i.e. the liquid is ideal) \eqref{KP} reduced to the Kadomtsev--Petviashvili equation \cite{Kadomtsev}. At $\gamma=0$ \eqref{KP} is the two--dimensional Korteweg--de Vries--Burgers equation \cite{Parkes1993}.

From relations \eqref{eq: non-dim_parameters} we see that the coefficient at the dispersion term in \eqref{KP} takes the form $\beta/2\alpha=(\rho_{l}R_{0}^{2}c_{0}^{2})/(6nP_{0}L)$ at $\sigma=0$ and one can see that the value of dispersion depends on the radius of bubbles, the polytrophic exponent and the pressure in the unperturbed state of the bubble--liquid mixture.

In the same way we can obtain that dissipation of the nonlinear waves depends on the liquid viscosity. In our consideration we do not take into account heat--transfer between the liquid and the gas in bubbles. However we can consider the influence of the heat--transfer on dissipation of nonlinear waves using the effective viscosity \cite{Nakor'akov,Nigmatulin}. This viscosity consist of the two parts: the liquid viscosity and the "thermal viscosity".

The "thermal viscosity" depends on the thermal conductivity of the gas, the volume gas content and the radius of bubbles in the unperturbed state,  the frequency of bubbles oscillations and the polytrophic exponent \cite{Nakor'akov,Nigmatulin}. For example, from the formula for effective viscosity \cite{Nakor'akov,Nigmatulin} we can see that in the case of a liquid with large viscosity and small bubbles the liquid viscosity causes dissipation of nonlinear waves. Otherwise dissipation of nonlinear waves is caused by heat--transfer and we can use the formula for the effective viscosity.



Replacing in \eqref{pre_main_main_dim_system_3} the liquid viscosity on the effective viscosity we can take into account the dissipation of nonlinear waves due to the heat--transfer.


\section{The Painleve test for the thee-dimensional nonlinear evolution equation.}

Using transformations
\begin{equation}
  \begin{gathered}
\tilde{\rho}_{1}=\frac{\beta}{2(\alpha+\alpha_{1})}\,u,\quad x=x',\quad t=\frac{2\alpha}{\beta}t' \vspace{0.1cm} \\
 y=\frac{1}{\delta}\sqrt{\frac{\beta}{2\alpha}}y', \quad z=\frac{1}{\gamma}\sqrt{\frac{\beta}{2\alpha}}z'
  \end{gathered}
\end{equation}
from \eqref{KP} we have the following equation (the primes are omitted)
\begin{equation}
\left(u_{t}+uu_{x}+u_{xxx}-\chi u_{xx}\right)_{x}+\frac{1}{2} \left(u_{yy}+u_{zz}\right)=0,
\label{rescaled_KP}
\end{equation}
where $\chi=\frac{\varkappa}{\beta}$.

Let us neglect the viscosity of a liquid.
In this case $\chi=0$ and \eqref{rescaled_KP} takes the form
\begin{equation}
\left(u_{t}+uu_{x}+u_{xxx}\right)_{x}+\frac{1}{2} \left(u_{yy}+u_{zz}\right)=0.
\label{KP_1}
\end{equation}

To investigate the integrability of \eqref{KP_1} let us apply the WTC-test with the Kruskal ansatz \cite{Kruskal1982,Hone2005}. We are looking for solution of \eqref{KP_1} in the form
\begin{equation}
u(x,y,z,t)=\Phi^{p}\sum\limits_{j=0}^{\infty}\,A_{j}\,\Phi^{j}, \quad A_{j}=A_{j}(y,z,t),\quad \Phi=x+\psi(y,z,t),
\label{main_expansion}
\end{equation}
where $\psi(y,z,t)$ is an arbitrary function.

The leading terms in \eqref{KP_1} are $u u_{xx}$, $u_{x}^{2}$, $u_{xxxx}$. Substituting \eqref{main_expansion} into these terms and equating coefficients at the lowest order of  $x+\psi(y,z,t)$ to zero we have
\begin{equation}
p=-2,\quad u_{0}=-12
\end{equation}
Substituting the expression
\begin{equation}
u=-12(x+\psi(y,z,t))^{-2}+A_{j}(x+\psi(y,z,t))^{j-2}
\end{equation}
into leading terms of \eqref{KP_1} and equating coefficient at $A_{j}$ to zero we have the following Fuchs indices
\begin{equation}
j_{1}=-1,\,j_{2}=4,\,j_{3}=5,\,j_{4}=6.
\end{equation}

We recall that \eqref {KP_1} has the Painlev\'{e} property if functions $ \psi (y, z, t) $, $A _ {4} (y, z, t) $, $A_{5}(y, z, t) $, $A_{6}(y, z, t) $ in \eqref{main_expansion} are arbitrary.

Substituting \eqref{main_expansion} into \eqref{KP_1} and consistently equating terms at various powers of $x +\psi (y, z, t) $ to zero we find
\begin{equation}
A_{1}(y,z,t)=0, \quad A_{2}(y,z,t)=-\psi_{t}-\frac{1}{2}(\psi_{y}^{2}+\psi_{z}^{2}), \quad
A_{3}=\frac{1}{2}\left(\psi_{yy}+\psi_{zz}\right).
\end{equation}
Functions $A_{4}(y,z,t)$, $A_{5}(y,z,t)$ can be taken as arbitrary. However the function $A_{6}(y,z,t)$ can be arbitrary if the following condition is satisfied
\begin{equation}
\psi_{yy}\,\psi_{zz}-\psi_{yz}=0.
\label{compatibility_condition}
\end{equation}

From \eqref{compatibility_condition} we see that  the function $ \psi (y, z, t) $ cannot be taken as arbitrary. Thus \eqref{KP_1} does not have the Painlev\'{e} property.

In the same way we can also show that \eqref{rescaled_KP} does not have the Painlev\'{e} property.

\section{Exact solutions of the three-dimensional nonlinear evolution equation}

We have obtained that \eqref{rescaled_KP} is not integrable in the general case and the Cauchy problem for this equation cannot be solved by the inverse scattering transform. However there are exact solutions of this equation. Let us construct some of them.

Using the traveling wave variables $u(x,y,z,t)=U(\theta),\,\theta=x+y+z-C_{0}\,t$ we have the nonlinear ordinary differential equation from \eqref{rescaled_KP}
\begin{equation}
C_{2}+C_{1}\,\theta+(1-C_{0})\,U+\frac{1}{2}\,U^{2}+U_{\theta\theta}-\chi U_{\theta}=0
\label{rescaled_KP_reduction}
\end{equation}
In the general case \eqref{rescaled_KP_reduction} cannot be solved but at $C_{1}=0$ \eqref{rescaled_KP_reduction}  is equivalent  to the traveling wave reduction of the Korteweg--de Vries--Burgers equation.  This equation was studied many times \cite{Parkes1993,Parkes96, Polyanin05, Kudryashov1988, Kudryashov1991}. As this takes place different methods can be applied  for finding exact solutions \cite{Biswas01, Kudr2010b, Vitanov01, Kudr2010dd, Kudr2010ee}. The general solutions of \eqref{rescaled_KP_reduction} at $C_{1}=0$ was obtained in \cite{Kudryashov2009}. It takes the form
\begin{equation}
U=C_{0}^{(1,2)}-1+\frac{6\chi^{2}}{25}-\exp\left\{\frac{2\chi \theta}{5}\right\}\wp\left(C_{3}-\frac{5}{\sqrt{12}\chi}\,
\exp\left\{\frac{\chi\theta}{5}\right\},0,C_{4}\right),
\label{rescaled_KP_reduction_elliptic_solution}
\end{equation}
where $\wp$ is the Weierstrass elliptic function, $C_{3},\,\,C_{4}$ are constants of integration. There is correlation on $C_{0}$ in the form
\begin{equation}
C_{0}^{(1,2)}=1\pm\sqrt{2\,C_{2}+\frac{36\chi^{4}}{625}}
\label{omega_def}
\end{equation}

\begin{figure}[h]
\begin{center}
 \includegraphics[width=140mm,height=100mm]{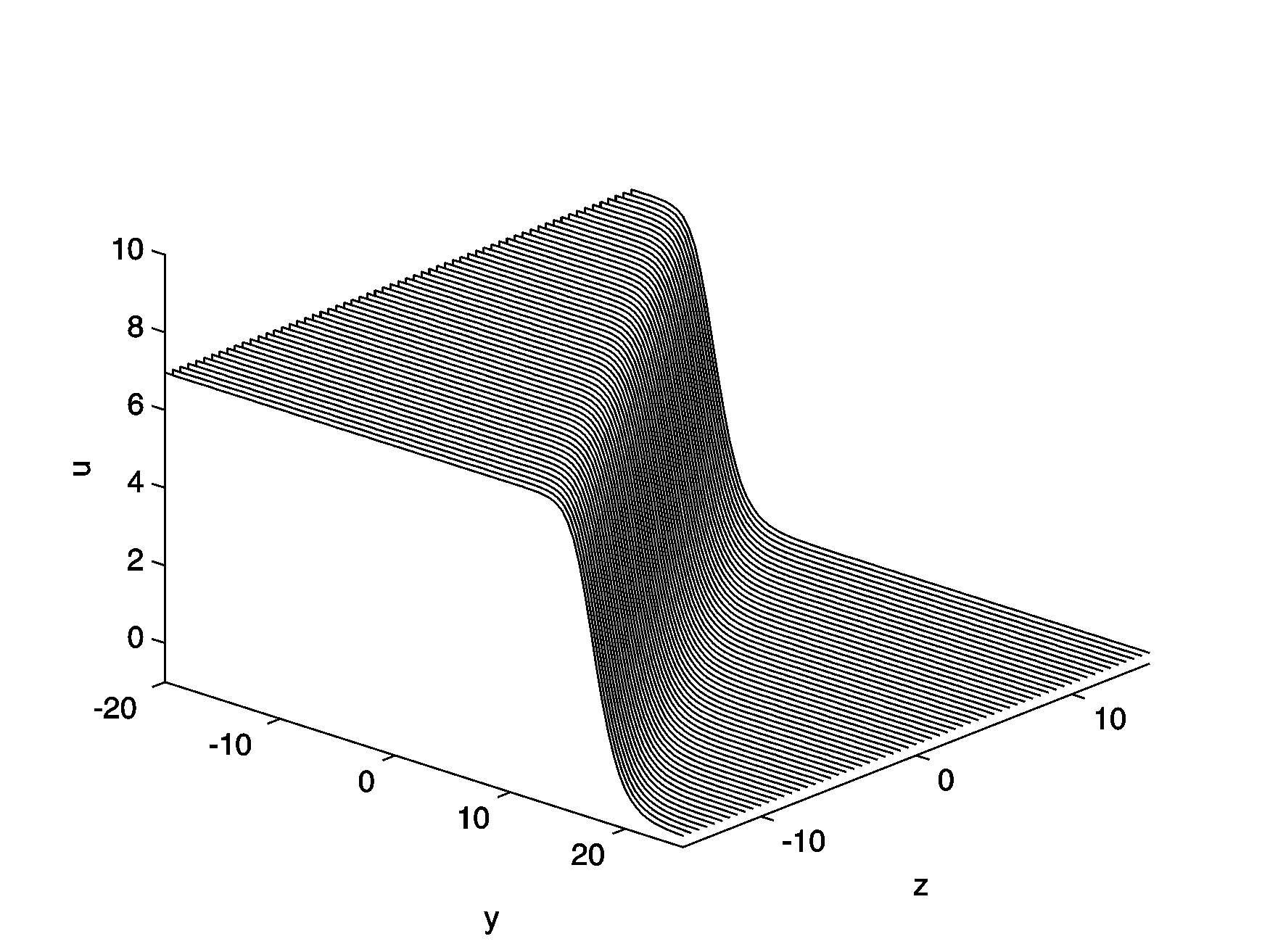}
 \caption{Solution \eqref{kink_solution} of equation \eqref{rescaled_KP} at $x=t=0$ }
\end{center}
 \label{Kink}
\end{figure}

From \eqref{rescaled_KP_reduction_elliptic_solution} at $C_{4}=0$ we obtain solitary wave solution of \eqref{rescaled_KP_reduction}  \cite{Kudryashov2009}
\begin{equation}
U=C_{0}^{(1,2)}-1+\frac{6\chi^{2}}{25}-\frac{\exp\left\{\frac{2\chi \theta}{5}\right\}}{\left(C_{3}\pm\frac{5}{\sqrt{12}\chi}\,\exp\left\{\frac{\chi \theta}{5}\right\}\right)^{2}},
\label{kink_solution}
\end{equation}
where $C_{0}^{(1,2)}$ is defined by \eqref{omega_def} as well. It is worth to note that this solution was obtained for the two--dimensional Korteweg--de Vries--Burgers equation in \cite{Parkes1993}.

We see that \eqref{kink_solution} describe the weak shock wave. The amplitude and steepness of this wave defined by parameter $\chi$ which depends on the liquid viscosity. Waves of type \eqref{kink_solution} in the one--dimensional case were obtained as solutions of Korteweg--de Vries--Burgers equation in \cite{Nakor'akov_1972,Nakor'akov1978}. Weak shock waves of type \eqref{kink_solution} were observed experimentally as well \cite{Wijngaarden1972,Nakor'akov1978,Nigmatulin}. The plot of solution \eqref{kink_solution} at $x=0$ and $t=0$ is demonstrated in Figure 1.

Let us consider equation \eqref{rescaled_KP}  at $ \chi=0$. In this case we neglect the liquid viscosity. As we show above equation \eqref{rescaled_KP} is not integrable at $\chi=0$. However we can construct multi--soliton solutions for this equation using special chose of coordinates. Following the Hirota bilinear method we look for the one--soliton solution of \eqref{rescaled_KP} in the form \cite{Ablowitz,Kudryashov_book}
\begin{equation}
u=A\,\frac{\partial^{2}}{\partial x^{2}} \ln F
\label{KP_soliton}
\end{equation}
where $F$ is the following
\begin{equation}
F=1+\exp\{k_{x}\,x+k_{y}\,y+k_{z}\,z-\Omega\,t+\phi_{0}\}, \quad \Omega=\frac{1}{k_{x}}\left[k_{x}^{4}+\frac{1}{2}\left(k_{y}^{2}+k_{z}^{2}\right)\right]
\label{KP_one_soliton}
\end{equation}
It is worth to note that $\Omega$  is defined by dispersion relation for \eqref{rescaled_KP}.

Substituting \eqref{KP_soliton} and \eqref{KP_one_soliton} into \eqref{rescaled_KP} we find that $A=12$. Thus one-soliton solution of \eqref{rescaled_KP} has the following form
\begin{equation}
\begin{gathered}
u=12\,\frac{\partial^{2}}{\partial x^{2}} \ln \Big( 1+\exp\Big\{k_{x}\,x+k_{y}\,y+k_{z}\,z -\frac{1}{k_{x}}\left[k_{x}^{4}+\frac{1}{2}\left(k_{y}^{2}+k_{z}^{2}\right)\right]\,t+\phi_{0}\Big\}\Big)
\label{KP_one_soliton_solution}
  \end{gathered}
\end{equation}

As we well know weakly nonlinear waves in bubbly liquids can be described by the Korteweg--de Vries equation \cite{Wijngaarden,Nakor'akov_1972,Wijngaarden1995}. We can consider solution \eqref{KP_one_soliton_solution} as three--dimensional counterpart for the one--soliton solution of the Korteweg--de Vries equation. Further we shall show that the one--dimensional one--soliton solution is stable to transverse perturbations.

We can find the two--soliton solution of \eqref{rescaled_KP} using formalism \eqref{KP_soliton} as well. In this case the function $F$ takes the following form \cite{Ablowitz,Kudryashov_book}
\begin{equation}
\begin{gathered}
F=1+\exp\{k_{x}^{(1)}\,x+k_{y}^{(1)}\,y+k_{z}^{(1)}\,z-\Omega_{1}\,t+\phi_{0}^{(1)}\}+ \\+\exp\{k_{x}^{(2)}\,x+k_{y}^{(2)}\,y+k_{z}^{(2)}\,z-\Omega_{2}\,t+\phi_{0}^{(2)}\}
+\exp\left\{(k_{x}^{(1)}+k_{x}^{(2)})\,x+\right.  \\ \left.+(k_{y}^{(1)}+k_{y}^{(2)})\,y+(k_{z}^{(1)}+k_{z}^{(2)})\,z-(\Omega_{1}+
\Omega_{2})\,t+\phi_{0}^{(1)}+\phi_{0}^{(2)}+\phi_{0}^{(1,2)}\right\}
\label{KP_two_soliton}
  \end{gathered}
\end{equation}

where
\begin{equation}
\Omega_{i}=\frac{1}{k_{x}^{(i)}}\left[(k_{x}^{(i)})^{4}+\frac{1}{2}\left((k_{y}^{(i)})^{2}+
(k_{z}^{(i)})^{2}\right)\right],\quad i=1,2
\end{equation}

\begin{figure}[h]
\begin{center}
 \includegraphics[width=140mm,height=100mm]{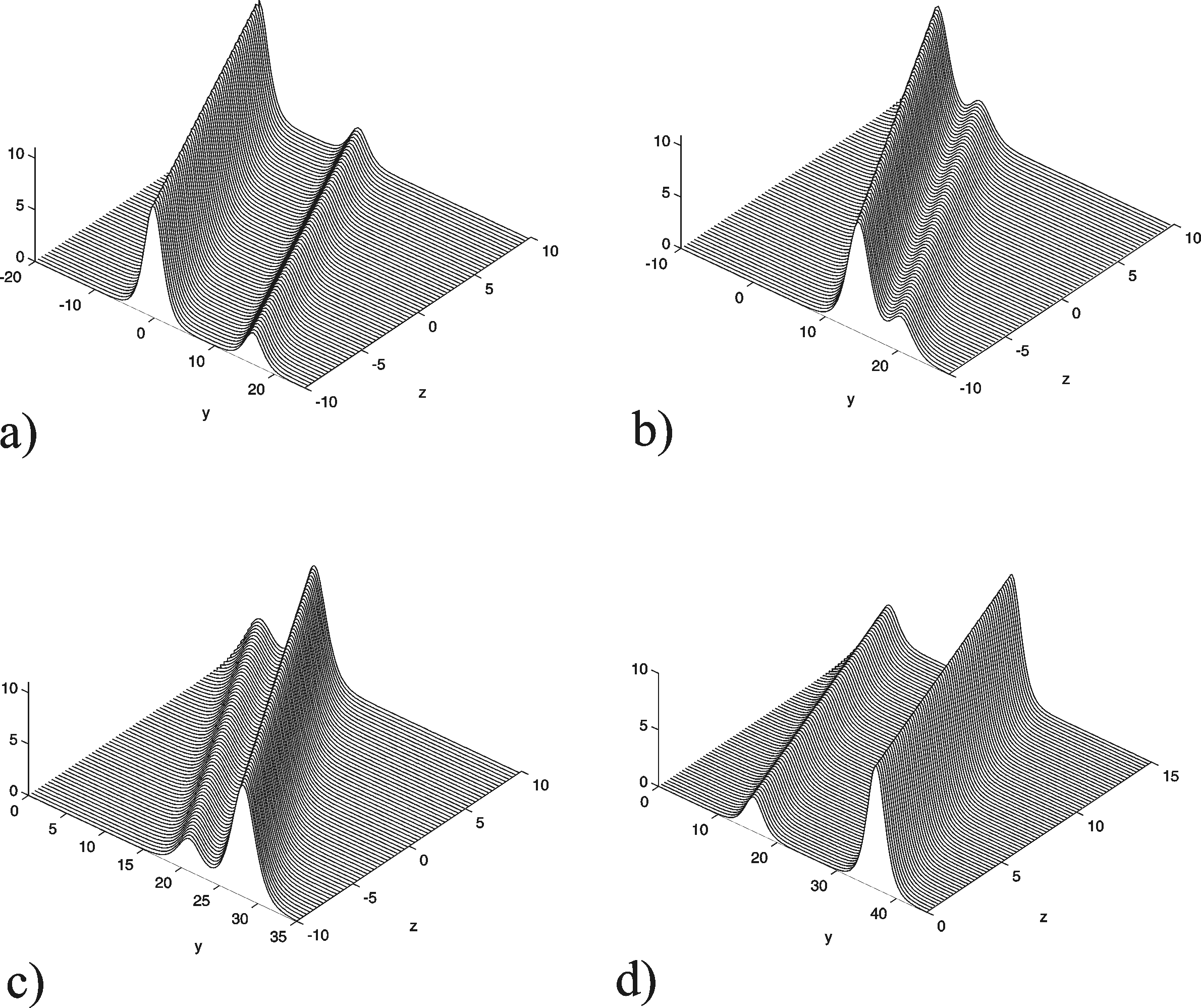}
 \caption{Two-soliton solution of equation \eqref{rescaled_KP} at different time moments a) $t=2$; b) $t=3.5$; c) $t=5.5$; d) $t=8$. }
\end{center}
 \label{Solitary_interaction}
\end{figure}

Substituting \eqref{KP_soliton} and \eqref{KP_two_soliton} into \eqref{rescaled_KP} we find that $A=12$ and $\phi_{0}^{(1,2)}$ is determined by the following relation
\begin{equation}
\begin{gathered} e^{\phi_{0}^{(1,2)}}=\left[6\,\left(k_{x}^{(1)}\right)^{2}\,\left(k_{x}^{(2)}\right)^{2}\left(k_{x}^{(1)}-k_{x}^{(2)}\right)^{2}-
  \left(k_{x}^{(1)}k_{y}^{(2)}-k_{x}^{(2)}k_{y}^{(1)}\right)^{2}-\right.\\ \left.-
  \left(k_{x}^{(1)}k_{z}^{(2)}-k_{x}^{(2)}k_{z}^{(1)}\right)^{2}\right] \left[
  6\,\left(k_{x}^{(1)}\right)^{2}\,\left(k_{x}^{(2)}\right)^{2}\left(k_{x}^{(1)}+k_{x}^{(2)}\right)^{2}
  -\right.\\ \left.-\left(k_{x}^{(1)}k_{y}^{(2)}-k_{x}^{(2)}k_{y}^{(1)}\right)^{2}-
  \left(k_{x}^{(1)}k_{z}^{(2)}-k_{x}^{(2)}k_{z}^{(1)}\right)^{2} \right]
  \label{phase_shift}
  \end{gathered}
\end{equation}

Substituting \eqref{KP_two_soliton} and \eqref{phase_shift} into \eqref{KP_soliton} at $A=12$ we have the two-soliton solution of \eqref{rescaled_KP}. This solution is the thee-dimensional counterpart for the two--soliton solution of the Korteweg--de Vries equation.

The plot of the two-soliton solution of \eqref{rescaled_KP} at $x=0$ at various moments of time is presented in Figure 2. From Figure 2 we can see that solitary waves described by \eqref{rescaled_KP} interact elastically.

The stability of the solitary wave solution for \eqref{KP} at $\gamma=0$ and $\varkappa=0$ (i.e. in a two-dimensional case) was shown in \cite{Kako1976}.
The full analysis of a stability of the solitary wave solution for \eqref{KP} in linear approach is given in \cite {Alexander1997} at $ \gamma=0$. It is shown, that  this solution is stable at $\delta> 0$. Let us show, that the results of \cite {Alexander1997} can be extended on \eqref{KP} at $ \gamma\neq0$.

Without loss of generality we can transform \eqref{KP} at $\varkappa=0$ to the following form
\begin{equation}
\left(4\tilde{u}_{t}-4\tilde{u}_{x}+4\tilde{u}\tilde{u}_{x}+
\tilde{u}_{xxx}\right)_{x}+3\tilde{u}_{yy}+3\tilde{u}_{zz}=0
\label{KP_Stability}
\end{equation}

We see that \eqref{KP_Stability} has the stationary solitary wave solution in the form
\begin{equation}
\tilde{u}^{(0)}=3\cosh^{-2}(x)
\end{equation}
Following \cite{Alexander1997} we consider small perturbations of the  solitary wave solution
\begin{equation}
\tilde{u}=\tilde{u}^{(0)}+e^{\lambda\,t+i\,k_{y}y+i\,k_{z}z}v(x)
\label{pertubed_solution}
\end{equation}
Substituting \eqref{pertubed_solution} into \eqref{KP_Stability} and  linearizing the result we have
\begin{equation}
v_{xxxx}-4[(1-\tilde{u}^{(0)})\,v]_{xx}+4\lambda\,v_{x}+3\,(k_{y}^{2}+k_{z}^{2})\,v=0.
\label{spectrum_equation}
\end{equation}

The spectrum of \eqref{spectrum_equation} and it's solutions were investigated in \cite{Alexander1997} assuming that $\lambda$ is a complex number.
It was shown that at real $k_{y}^{2}+k_{z}^{2}>0$ solutions of \eqref{spectrum_equation} are bounded. Hence for real $k_{y} $ and $k_ {z}$ solitary wave solutions governed by the equation \eqref{KP} at $ \varkappa=0$ are stable.

\section{Conclusion}

Let us formulate the results of this work shortly. The one--dimensional nonlinear evolution equations for the description of waves in a liquid containing gas bubbles are well known but for the best of our knowledge there are no nonlinear evolution equations for description of nonlinear waves in a liquid containing gas bubbles in the three-dimensional case. In this work` we have considered the three--dimensional nonlinear waves in a bubble--liquid mixture taking into account the viscosity of liquid. We have obtained that the waves can be described by the generalization of the Korteweg--de Vries--Burgers equation and the Kadomtsev--Petviashvily equation.

We have checked this nonlinear evolution equation on the Painlev\'e property using the WTC test with the Kruskal anzats. We have found that \eqref{rescaled_KP}  is not integrable in the general case. However there are some exact solutions of \eqref{rescaled_KP} for description of nonlinear waves in a liquid with gas bubbles. We have presented traveling wave solutions for this equation.
In the case of 	non-viscous liquid the nonlinear evolution equation is nonintegrable as well. But we have obtained one--soliton and multi--soliton solutions for this equation.
We have shown that solitary waves governed by \eqref{rescaled_KP} in the absence of the dissipative losses is stable in relation to transverse perturbations.


\end{document}